\documentclass{article}
\usepackage{amssymb}
\usepackage{graphicx} 
\usepackage{tikz}     
\usepackage{url}      
\usepackage{amsmath}
\usepackage{titling}
\usepackage{authblk}  

\usepackage[margin=1.5in]{geometry}

\newtheorem{example}{Example}

\newtheorem{definition}{Definition}

\usepackage{placeins} 
\usepackage{float}    
\title{Research Vision: Multi-Agent Path Planning for Cops And Robbers Via Reactive Synthesis}
\author[1]{William Fishell}
\author[2]{Andoni Rodríguez}
\author[3]{Mark Santolucito}
\affil[1]{Columbia University, USA. william@example.com}
\affil[2]{IMDEA Software Institute, Spain. andoni.rodriguez@imdea.org}
\affil[3]{Barnard College, Columbia University, USA. msantolu@barnard.edu}

\date{March 13, 2025}

\begin{document}
\setlength{\droptitle}{-3cm}
\maketitle

\begin{abstract}
    We propose the problem of multi-agent path planning for a generalization of the classic Cops and Robbers game via reactive synthesis. Specifically, through the application of LTL\(t\) and Coordination Synthesis, we aim to check whether various Cops and Robbers games are realizable (a strategy exists for the cops which guarantees they catch the robbers). Additionally, we construct this strategy as an executable program for the multiple system players in our games.
    In this paper we formalize the problem space, and propose potential directions for solutions.
    We also show how our formalization of this generalized cops and robbers game can be mapped to a broad range of other problems in the reactive program synthesis space. 
\end{abstract}

\section{Introduction}
Reactive synthesis is classically modeled as a game, though often applied to domains such as arbiter circuits and communication protocols~\cite{finkbeiner2016synthesis}. 
We aim to show how reactive synthesis can be applied to a literal game - cops and robbers - to generate strategies for agents in the game.
We propose a game that requires the coordination of multiple agents in a space of datatypes and operations that are richer than is easily captured by the traditional Linear Temporal Logic (LTL) approach of synthesis over Boolean streams~\cite{pnueli1977temporal}.  
In particular, we draw inspiration from prior work on Coordination Synthesis~\cite{coordination_synthesis}, LTL moduluo theories (LTL\(t\))~\cite{Reactive_Synthesis_Refs}, and Temporal Stream Logic Moduluo theories (TSL-MT)~\cite{finkbeiner2019temporal, choi2022can} to describe our problem and potential solution spaces. 
The traditional game~\cite{nowakowski1983vertex} asks whether \(K\) cops can catch a single robber on a graph.
In a temporal logic setting, this amounts to a safety condition on the robbers (they are never caught by the cops), and the dual liveness condition for the cops (they eventually catch the robbers).
We modify the traditional graph theory focused version of the game to have a more visual game on a grid system, allowing for various configurations, including:
\begin{itemize}
    \item An environment with various node types such as walls, safe zones, and open spaces. 
    \item Potentially multiple robbers, with a both global survival and individual survival constraints.
    \item Robbers aim to satisfy a liveness constraint by alternating between safe zones.
\end{itemize}

One approach to this problem is build an LTL encoding of both the arena and the properties.
However, this is unlikely to scale as the grid grows in size - the number of possible states in a direct encoding for a grid of NxM with C cops and R robbers is $(N+M)^{C+R}$.
For a simple 10x10 grid with 2 cops and 1 robber, this is 1,000,000 states - far outside the scope of any current LTL synthesis engine~\cite{syntcomp}.
We hypothesize that, using LTL\(t\), we simplify the state space and construct reactive programs that enable coordination synthesis for verifying game realizability. 
Next, we plan to use the constructed controller to synthesize a strategy for the agent players as they move throughout the environment. 
These modifications extend the traditional game into more complex settings applicable to path planning and task execution.
We address the problem through a combination of coordination synthesis, LTL\(t\), and Boolean abstraction \cite{Boolean_Abstractions}.

In summary, the key contributions of this research vision paper are:
\begin{enumerate}
    \item The formalization of Cops and Robbers through the lens of temporal logic and reactive synthesis.
    \item A collection of variations of this game, showing the flexibility of the problem definition to add various challenges to the synthesis problem.
    \item An outline of one approach to solving this problem using a combination of LTL\(t\) and Coordination Synthesis.
\end{enumerate}

\begin{figure}[H]
\centering
\includegraphics[width=0.6\linewidth]{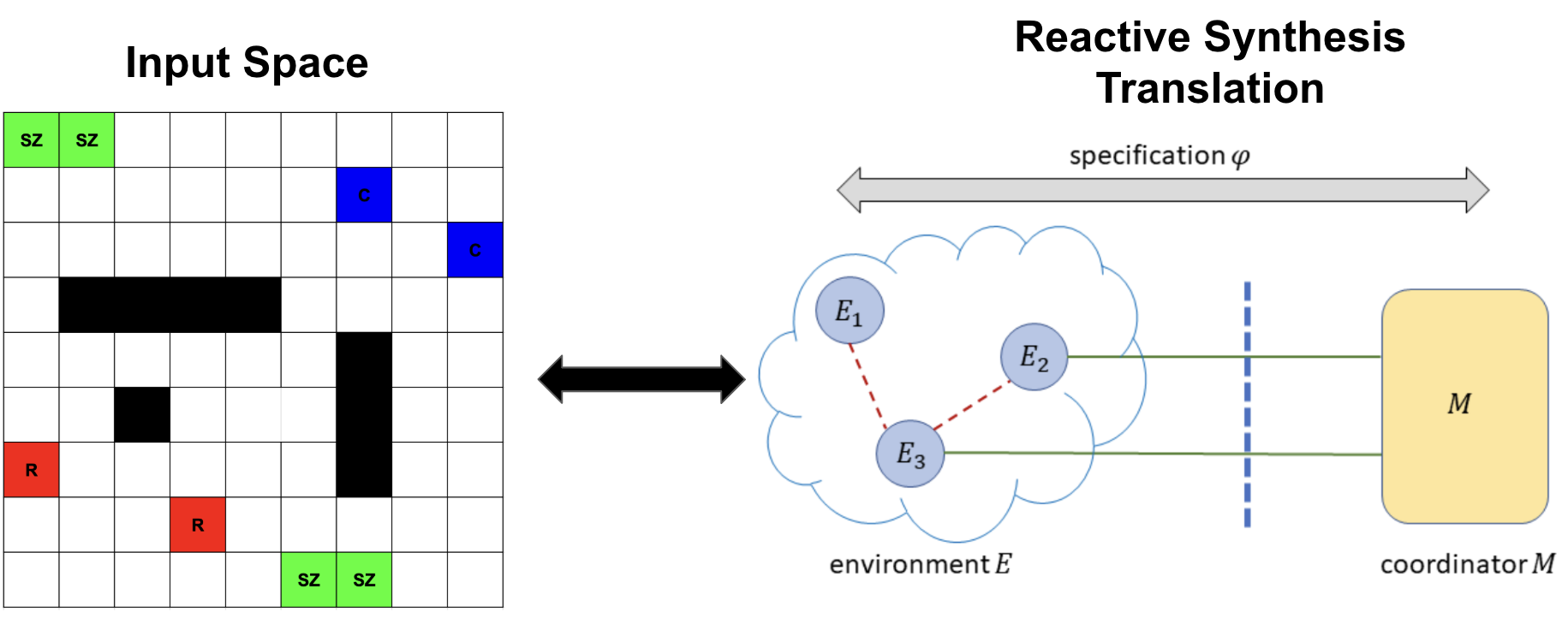}
\caption{Problem Visualization and Translation}
\end{figure}

\textbf{Related work} \\
The problem of Cops and Robbers spans multiple domains, including reinforcement learning, robotics, computer vision, graph theory, and formal methods. Recent research has focused on deep reinforcement learning for multi-agent Cops and Robbers systems, with applications in robotics \cite{MultiRobotPusuitEvasion,xu2024approximate}. Meanwhile, graph theory has extensively investigated various graph constructions and their impact on the realizability of Cops and Robbers \cite{crawford2024k,erlebach2024cop}.
\\
\\
In contrast, our approach applies reactive synthesis to check the realizability of certain initial states and environments and to synthesize a winning strategy. This method aims to achieve coordination similar to that of reinforcement learning models but with formal guarantees that the resulting strategy is correct. Additionally, like graph theory studies that explore a wide array of graphs, our setup is designed to be flexible enough to analyze a broad range of graph structures. Exploring this problem through reactive synthesis will yield valuable results not only for coordination synthesis and the Cops and Robbers game but also for reinforcement learning and graph theory.

\section{Preliminaries}

\subsection{LTL and Reactive Synthesis}

We start from LTL~\cite{pnueli1977temporal}, which has the following syntax:
\[
  \varphi  ::= \top | a | \varphi \lor \varphi | \neg \varphi
  | X \varphi | \varphi \mathcal{U} \varphi, 
\]
where  $a\in AP$ is an \emph{atomic proposition}, 
$\{\land,\neg\}$ are the common Boolean operators of \emph{conjunction} and
\emph{negation}, respectively, and
$\{X,\mathcal{U}\}$ are the \emph{next} and \emph{until} temporal
operators, respectively.
Additional temporal operators include  $F$(\emph{finally}), and
 $G$ (\emph{always}), which can be derived from the syntax above.
Given a set of atomic propositions $\overline{a}$ we use $vals(\overline{a})$ for a set of possible valuations of variables in $\overline{a}$ (i.e. $vals(\overline{a})=2^{\overline{a}}$), and we use $v_{\overline{a}}$ to range over $vals(\overline{a})$.
We use $\Sigma=vals(AP)$.
The semantics of LTL formulas associates traces $\sigma\in\Sigma^\omega$ with
LTL fomulas (where $\sigma \models \top$ always holds, and $\vee$ and $\neg$ are standard):
\[
  \begin{array}{l@{\hspace{0.3em}}c@{\hspace{0.3em}}l}
    \sigma \models a & \text{iff } & a \in\sigma(0) \\
     \sigma \models X \varphi & \text{iff } & \sigma^1\models \varphi \\
     \sigma \models \varphi_1 \mathcal{U} \varphi_2 & \text{iff } & \text{for some } i\geq 0\;\; \sigma^i\models \varphi_2, \text{ and } \\
    && \;\;\;\;\;\text{for all } 
    \text{for all } 0\leq j<i, \sigma^j\models\varphi_1 \\
  \end{array}
\]

Reactive LTL synthesis~\cite{pnueli1977temporal}
is the task of producing a system that satisfies a given LTL specification $\varphi$, where
atomic propositions in $\varphi$ are split into variables
controlled by the environment (``input variables'') and by the system (``output variables''), denoted by $\overline{e}$ and $\overline{s}$, respectively.
Synthesis corresponds to a game where, in each turn, the
environment player produces values for the input propositions, and the system player
responds with values of the output propositions.
A play is an infinite sequence of turns, i.e., an infinite interaction of the system with the environment.
%
%
A strategy  for the system is a tuple $\rho: \langle{Q,q_0,\delta,o}\rangle$
where $Q$ is a finite set of states, $q_0\in Q$ is the inital state,
$\delta:Q\times vals(\overline{e}) \rightarrow Q$ is the transition function and
$o:Q\times vals(\overline{e}) \rightarrow vals (\overline{s})$ is the output function.
$\rho$ is said to be \emph{winning}
for the system if all the possible plays played
according to the strategy satisfy the LTL formula $\varphi$.
Realizability is the decision problem of whether there is a winning strategy for the system , 
while LTL synthesis is the computational problem of producing one. 

\subsection{Synthesis Modulo Theories}

Recently, more expressive temporal specification  logics have been proposed for reactive synthesis \cite{katis2018validity,samuel21gensys}.
The most popular among them is Temporal Stream Logic (TSL) \cite{finkbeiner2019temporal}, which extends linear-time temporal logic with uninterpreted predicates and functions. Moreover, TSL Modulo Theories (TSL-MT) \cite{finkbeiner21temporal,choi2022can,maderbacher2021reactive,heim24solving,schmuck24localized,heim25efficient} generalizes TSL with first-order theories. TSL-MT specifications can refer to any arbitrary first-order theory, such as the theory of Linear Integer Arithmetic
(LIA) or the Theory of Arrays. TSL is the special case of TSLMT where all symbols are from the Theory of Uninterpreted Functions.
As we can see, TSL-MT is very expressive and has been followed by a lot of work, but its design follows the goal of easy extraction of programs (via a finite
number of possible assignments), which makes it difficult to specify more declarative specifications.

Therefore, in this research vision, we plan to use an alternative logic: LTL modulo theory (LTLt) \cite{geatti22linear,geatti23decidable,Boolean_Abstractions}, whose syntax replaces atoms $a$ by literals $l$ from some theory $\mathcal{T}$.
In particular, given an LTLt formula $\varphi(\overline{z})$ with variables $\overline{z}$ the semantics of LTLt now associate traces $\sigma$ (where each letter is a valuation of $\overline{z}$, i.e., a mapping from $\overline{z}$ into domain $\mathbb{D}$) with LTLt  formulae.
The semantics of the Boolean and temporal operators are as in LTL, and for literals it holds:
\[
  \begin{array}{l@{\hspace{0.3em}}c@{\hspace{0.3em}}l@{\hspace{0em}}}
    \sigma \models l & \text{iff } & \text{the valuation $\sigma(0)$ of $\overline{z}$ makes $l$ true according to $\mathcal{T}$} \\
  \end{array}
\]

The synthesis problem for LTLt \cite{Boolean_Abstractions,Reactive_Synthesis_Refs,rodriguez24realizability,rodriguez24adaptive,rodriguez25shield,corsi24verification} is analogous to the one in LTL, but the strategy is a (symbolic) representation of an infinite-state machine and all variables $\overline{e}$ and $\overline{s}$ belong to the domain $\mathbb{D}$ of $\mathcal{T}$.


\newpage

\section{Traditional Cops And Robbers}

Cops and Robbers is a classic graph theory game where one or more cops and a robber take turns moving along adjacent nodes on a graph, with the cops aiming to capture the robber by occupying the same node. Formally:

\begin{definition} \label{def:traditional}
Let us define an arena as a graph $A=\{V,E\}$, where V is a set of vertices, and E is a set of edges such that $E \subseteq \{(u,v) | u, v \in V\}$.
Also, we are given initial positions $I_c \subseteq V = \{i_c^0, i_c^1, ...i_c^{C-1}\}$ of $C$ cops (where for all $elemt1, elemt2 \in I_c)$ it is never the case $elemt1 = elemt2$).
Also, initial position $I_r  \subseteq V =\{i_r\}$ of $R=1$ robber such that $I_c \cap  I_r = \emptyset$.
Then, we denote the position of each cop in a given timestep $t$ with $P_c \subseteq V = \{p_c^{0,t}, p_c^{1,t}, ...p_c^{{C-1},t}\}$ and the position of the robber with $P_r  \subseteq V =\{p_r^t\}$.

For any cop \(P_c\) with index \(J\), if at time \(t\) the cop is at position \(P_c^{J,t}\), then for time \(t+1\) the cop must move to a new position \(P_c^{J',t+1}\) such that: $P_c^{J',t+1} \neq P_c^{J,t}$. 
Similarly, for any robber \(P_r\) with index \(q\), if at time \(t\) the robber is at position \(P_r^{q,t}\), then for time \(t+1\) the robber must move to a new position \(P_r^{q`,t+1}\) such that: $P_r^{q`,t+1} \neq P_r^{q,t}.$
Lastly, players of the same team cannot switch positions during their turns. Without loss of generality, consider two cops $P_c^i$ and $P_c^j$ at positions $u$ and $v$ at time $t$ such that ${(u,v)}\in E$. If at time $t+1$ $P_c^i$'s position is $v$ then $P_c^j$'s position cannot be $u$.

The realizability problem asks: is there a reachability strategy $\rho$ for the cops such that at some time $t \geq 0$, at least one of $p_c^{j,t} \in P_c$ is in a position such that $p_c^{j,t}=p_r^t$? The cooresponding synthesis problem asks to then construct such a strategy $\rho$. 
\end{definition}

Note how this problem can be encoded as a synthesis of reactive systems problem, where the environment (cops) and the system (robber) alternate turns moving on $A$ and either of them wins (i.e., $\rho$ exists or does not exist). Consider the LTL specification
\begin{equation}
\begin{aligned}
\phi =\; & \mathbf{G}\,\neg \text{CopsCollide} \land \mathbf{G}\,\neg \text{RobbersCollide} \land \mathbf{G}\,\neg \text{RobbersSwitch} \\
         & \land \mathbf{G}\,\neg \text{CopsSwitch} \land \mathbf{G}\,\neg \text{RobberCaught} \land \mathbf{G}\,(\text{CopsMove} \land \text{RobberMove})\\
         & \land \mathbf{G}\,(\text{Move}\to \mathbf{X}\,\text{NewVertex})
\end{aligned}
\label{eq:phi}
\end{equation}

\begin{example}
    
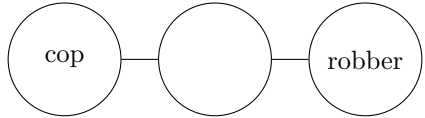
\begin{figure}[H] \label{fig:losing}
  \caption{Losing Cops \& Robbers Game: \textbf{Robber System Player}}
  \vspace{1em}
  \centering
  \begin{tikzpicture}[node distance=2cm,
    every node/.style={circle, draw, minimum size=1.5cm, inner sep=0pt, align=center}]
    \node (cop) {cop};
    \node[right of=cop] (middle) {};
    \node[right of=middle] (robber) {robber};

    \draw (cop) -- (middle) -- (robber);
  \end{tikzpicture}
\end{figure}

Given the simple LTL specification~$\eqref{eq:phi}$,
the system defined in Fig.~\ref{fig:losing} is unrealizable (i.e., $\rho$ exists). 
\end{example}

When the environment grows (i.e., $|V|$ increases) and the single robber assumption is dropped (i.e., $|P_r|>1$), how can we ensure that the robbers remain safe by construction? There is comparatively less research on how increasing $|P_r|$, the number of robbers, affects the system, and this modification introduces complexities.

\begin{figure}[H] \label{fig:crowding}
  \centering
  \begin{tabular}{cc}
    \begin{tikzpicture}[every node/.style={circle, draw, minimum size=1.5cm, inner sep=0pt, align=center}]
      \node (cop) at (0,0) {cop};
      \node (robber) at (4,0) {robber};
      
      \node (empty1) at (2,1) {};
      \node (empty2) at (2,-1) {};
      
      \draw (cop) -- (empty1);
      \draw (cop) -- (empty2);
      \draw (robber) -- (empty1);
      \draw (robber) -- (empty2);
    \end{tikzpicture}
    &
    \begin{tikzpicture}[every node/.style={circle, draw, minimum size=1.5cm, inner sep=0pt, align=center}]
      \node (cop) at (0,0) {cop};
      \node (robber) at (4,0) {robber};
      
      \node (robber2) at (2,1) {robber};
      \node (empty) at (2,-1) {};
      
      \draw (cop) -- (robber2);
      \draw (cop) -- (empty);
      \draw (robber) -- (robber2);
      \draw (robber) -- (empty);
    \end{tikzpicture} \\
    (a) Winning Cops and Robbers Game & (b) Losing Cops and Robbers Game Due to Crowding \\
  \end{tabular}
  \caption{Effect of introducing more robbers into realizable situations.}
\end{figure}
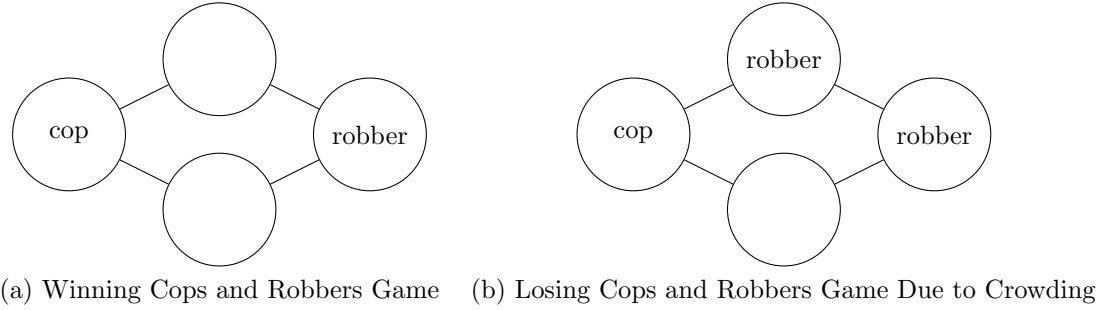

\FloatBarrier 

For instance, adding new system players (robbers) introduces crowding, which can make some previously realizable solutions unrealizable (see Fig.~\ref{fig:crowding}). 
Furthermore, the traditional Cops and Robbers game of Def.~\ref{def:traditional} has several limitations that make it unsuitable for understanding multi-agent adversarial systems:
\begin{itemize}
    \item \textbf{Lack of Liveness Constraints:} The pursuit-evasion dynamics focus solely on safety by moving an agent to ensure evasion, without addressing liveness properties that may be critical in complex scenarios.
    \item \textbf{Simplistic Environmental Modeling:} Nodes are modeled as merely occupied or unoccupied by players, which limits the representation of environmental complexity.
    \item \textbf{Limited Multi-Agent Exploration:} The traditional formulation has not been extensively studied for systems involving multiple players or adversaries.
\end{itemize}
We propose a series modified Cops and Robbers game that addresses these issues and provides a framework for verifying various multi-agent adversarial systems. 

\section{Generalized Cops And Robbers}

The most general version of Cops and Robbers is an imperfect-information evasion–pursuit game played on a  graph \(A\) where  \(|V|=\infty\) with \(C\) cops and \(R\) robbers, where players see only their adjacent neighbors. Although the objectives and movement rules remain unchanged to those in Def. 1, we employ Knowledge LTL~\cite{rutledge2023controller} (\(K\)LTL) to encode the knowledge constraints, yielding an LTL specification for the game. Our approach takes the baseline specification \(\Phi\) and defines a family of more restrictive variants by conjoining it with an extra constraint \(\psi\); that is, each modified game is expressed as \(\Phi \land \psi\).
\begin{equation}
\begin{aligned}
\Phi =\; & \mathbf{G}\,\neg \text{CopsCollide} \land \mathbf{G}\,\neg \text{RobbersCollide} \land \mathbf{G}\,\neg \text{RobbersSwitch} \\
         & \land \mathbf{G}\,\neg \text{CopsSwitch} \land \mathbf{G}\,\neg \text{RobberCaught} \land \mathbf{G}\,(\text{CopsMove} \land \text{RobberMove})\\[1mm]
         & \land \mathbf{G}\,(\text{Move}\to \mathbf{X}\,\text{NewVertex})\\[1mm]
         & \land \mathbf{G}\,\Bigl(\neg \text{Adjacent} \to \bigl(\neg K_{\text{Cop}}(\text{RobberPos}) \land \neg K_{\text{Robber}}(\text{CopPos})\bigr)\Bigr)
\end{aligned}
\label{master_imperfect}
\end{equation}
\textbf{Note: the analagous perfect information specification is the same in every other way. Thus, denoting perfect information games we denote it as $\Phi'\land \psi$ where $\Phi'$ is Formally:} 
\begin{equation}
\begin{aligned}
\Phi'=\Phi/K_{\text{Cop}}(\text{RobberPos}) \land \neg K_{\text{Robber}}(\text{CopPos})
\end{aligned}
\end{equation}

\section{Modified Cops And Robbers Game, Part I: \\ Robbers as System Players}
\textbf{Research Question}\\
How does the addition of a safe zone livness constraint change the synthesis problem?
\\
\\
\textbf{Hypothesis}
The problem becomes solvable because the safe zones function as rest nodes enabling system players to not remember a large history of previous moves.
\subsection{Problem Overview}
This modified version of cops and robbers is played on a graph with $S$ safe zones s.t. $S \ge2$ and each safe zone encompasses an arbitrary number of adjacent nodes.  In this variant, the robbers’ goal is to indefinitely alternate between the two safe zones—akin to moving back and forth in cricket—without being caught. Once a robber enters a safe zone, the cops cannot reach them. However, a robber must exit the safe zone and reenter the grid within $\le$ 2 turns. Furthermore, after exiting one safe zone, a robber must eventually visit the other safe zone and must do so before returning to the safe zone they just left.
\\
\\
Like all of our modified games going forward, this game is projected on an \(N\times N\) grid which satisfies all of the valid connections between nodes $u,v\in A$. The rest of the grid is filled in as walls which are immovable and cannot be occupied. In total, the game has 5 types of cells: \textit{Cops} (Blue), \textit{Robbers} (Red), \textit{Open Spaces} (White), \textit{Walls} (Black), and \textit{Safe Zones} (Green) with $C$,$R\in$ \(N\) cops and robbers respectively. 
\\ 
\\
We explore whether robbers can coordinate without crowding each other and if the controller can devise a strategy that ensures both survival and task completion in various imperfect and perfect information game settings on bounded and unbounded grids. 

\begin{figure}[H]
\centering
\includegraphics[width=0.3\linewidth]{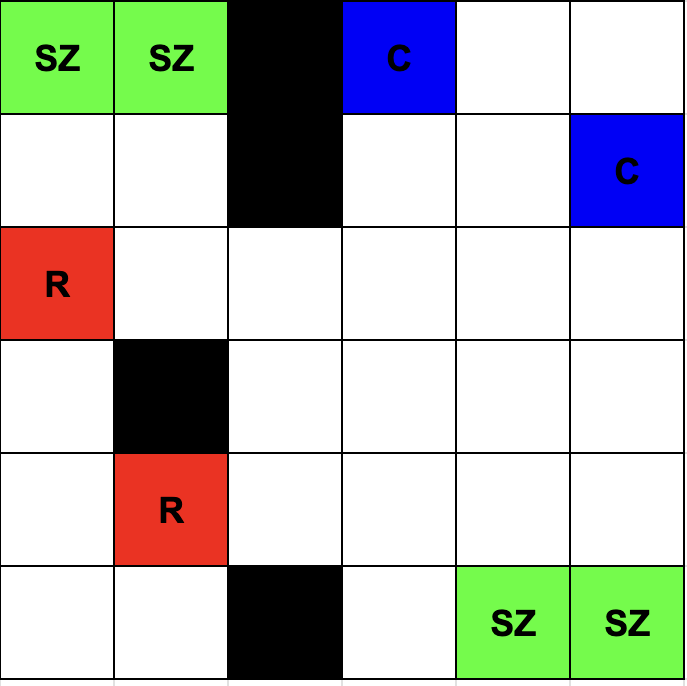}
\caption{Sample initial starting position for modified game: robbers as system players with 2 safe zones, each comprising 2 nodes}
\end{figure}
Now, we propose two variants of the game: a perfect information version and an imperfect information version.

\subsection{Perfect Information Game}
The perfect information game can be described by the LTL specification 
\begin{equation}
\begin{aligned}
\varphi =\; & \Phi' \\
& \land\, \mathbf{G}\mathbf{F}\,(\text{SafeZone}_1 \lor \text{SafeZone}_2) \\
& \land\, \mathbf{G}\Bigl(\text{SafeZone}_1 \to \mathbf{F}\,\text{SafeZone}_2\Bigr) \\
& \land\, \mathbf{G}\Bigl(\text{SafeZone}_2 \to \mathbf{F}\,\text{SafeZone}_1\Bigr) \\
& \land\, \mathbf{G}\Bigl(\text{SafeZone} \to \bigl(\mathbf{X}\,\neg \text{SafeZone} \lor \mathbf{X}\mathbf{X}\,\neg \text{SafeZone}\bigr)\Bigr) \\
& \land\, \mathbf{G}\Bigl((\text{SafeZone}_1 \land \mathbf{X}\,\neg \text{SafeZone}_1) 
         \to \bigl(\neg \text{SafeZone}_1\, \mathbf{U}\, \text{SafeZone}_2\bigr)\Bigr) \\
& \land\, \mathbf{G}\Bigl((\text{SafeZone}_2 \land \mathbf{X}\,\neg \text{SafeZone}_2) 
         \to \bigl(\neg \text{SafeZone}_2\, \mathbf{U}\, \text{SafeZone}_1\bigr)\Bigr) \\
& \land\, \mathbf{G}\,\neg \text{CollideWithWall}\,.
\end{aligned}
\label{eq:varphi}
\end{equation}

\subsubsection{Finite Grid}
In this game, every player knows the location of every other player and the location of all walls and safezones in the grid. The inputs to this system will be a game state at the time \(T(i)\). The outputs of this game will be updated positions for the robbers.
\subsubsection{Infinite Grid}
In the unbounded grid there is added complexity in ensuring \textit{perfect information} as it is impossible to input all information of an unbounded environment. Thus, the grid  will be described by sets of piecewise functions which are applied to determine the layout of walls and open spaces. The agents can then determine whether a position is a wall or a free space through a simple function call on the coordinate. 
\\
\textbf{For Example:}\\
We partition the integer lattice 
\[
\mathbb{Z}^2 = \{(x,y) : x,y \in \mathbb{Z}\}
\]
into two types of regions:

\begin{enumerate}
    \item \textbf{Checkerboard Region:} In areas where 
    \[
    (x+y) \bmod 100 < 25 \quad \text{or} \quad (x+y) \bmod 100 \ge 75,
    \]
    every square is colored according to a standard checkerboard rule:
    \[
    C(x,y) = \begin{cases}
    \text{white}, & \text{if } x+y\equiv 0 \pmod{2},\\[1mm]
    \text{black}, & \text{if } x+y\equiv 1 \pmod{2},
    \end{cases}
    \]
    
    \item \textbf{L‐Shape Region:} In the complementary band 
    \[
    25 \le (x+y) \bmod 100 < 75),
    \]
    the plane is tiled with a family of L‐shaped polyominoes.
\end{enumerate}

Since iterating through the entire infinite state space is impossible, we partition the game board into finite \(N \times N\) grids. Each grid is centered on a robber, and the configuration of walls within that grid is determined by the function call. Moreover, each robber not only knows the positions of safe zones, cops, and walls in their own grid but also shares this information with the other robbers, ensuring that every robber is aware of all bounded areas. Similarly to the finite grid the outputs of this system will be the updated positions for the robbers. Through this construction of the grid and partitioning of the infinite space, we will be able to capture the important aspects of a perfect information system. The major potential problem for this construction is a state space explosion. We will look into two main methods to avoid this:
\begin{itemize}
    \item \textbf{Keep the number of agents in the system small}
    \item \textbf{Take advantage of overlapping regions between system players to avoid duplication}
\end{itemize}
\subsection{Imperfect Information Game}
The imperfect information game can be described by the LTL specification 
\begin{equation}
\begin{aligned}
\varphi =\; & \Phi \\
& \land\, \mathbf{G}\mathbf{F}\,(\text{SafeZone}_1 \lor \text{SafeZone}_2) \\
& \land\, \mathbf{G}\Bigl(\text{SafeZone}_1 \to \mathbf{F}\,\text{SafeZone}_2\Bigr) \\
& \land\, \mathbf{G}\Bigl(\text{SafeZone}_2 \to \mathbf{F}\,\text{SafeZone}_1\Bigr) \\
& \land\, \mathbf{G}\Bigl(\text{SafeZone} \to \bigl(\mathbf{X}\,\neg \text{SafeZone} \lor \mathbf{X}\mathbf{X}\,\neg \text{SafeZone}\bigr)\Bigr) \\
& \land\, \mathbf{G}\Bigl((\text{SafeZone}_1 \land \mathbf{X}\,\neg \text{SafeZone}_1) 
         \to \bigl(\neg \text{SafeZone}_1\, \mathbf{U}\, \text{SafeZone}_2\bigr)\Bigr) \\
& \land\, \mathbf{G}\Bigl((\text{SafeZone}_2 \land \mathbf{X}\,\neg \text{SafeZone}_2) 
         \to \bigl(\neg \text{SafeZone}_2\, \mathbf{U}\, \text{SafeZone}_1\bigr)\Bigr) \\
& \land\, \mathbf{G}\,\neg \text{CollideWithWall}\,.
\end{aligned}
\label{eq:varphi2}
\end{equation}
\subsubsection{Finite Grid}
In the imperfect information game, each player only knows the locations of other players and walls within a Manhattan radius \(R\) of their current \((x,y)\) position--this area is called the Zone of Interest (\(ZI\)). All system players always know the safe zones. Players do not retain information from previously explored areas; each reactive program receives as input only the state space of the grid within its agent's \(ZI\). Similar to the perfect information game, the outputs of these reactive programs are the updated positions of the robbers.
\subsection{Infinite Grid}
The inputs for the infinite game are virtually identical to the finite game for imperfect information. There is no shared information and players are only aware of the other states in their \(ZI\). Although this is an imperfect information game, we need to construct the grid in a way such that every cell can be determined once it enters an agent's \(ZI\), so we rely on the same piece-wise approach used in the perfect information game to generate the walls and open spaces. 

\section{Modified Cops And Robbers Game, Part II: \\ Cops as System Players}
\textbf{Research Question}\\
How does the original evasion pursuit game change when we modify how system players interact with each other and their environments? \\
\\
\\
\textbf{Hypothesis}
By adding rules that let the cops (system players) acquire additional information, specific behaviors will be incentivized which appear counterintuitive to satisfying their liveness constraint.
\subsection{Problem Overview}
Assigning the role of system players to the cops requires us to modify $\Phi$ such that our safety constraint becomes a liveness constraint. Consider $\Pi$  defined as 
\begin{equation}
\begin{aligned}
\Pi=(\Phi/\mathbf{G}\,\neg \text{RobberCaught})\land\mathbf{F}\, \text{RobberCaught}
\end{aligned}
\end{equation}
for our modified game. \\ 
\textbf{Note: this new constructions is just for simpler convention and is logically equivalent to the analogous $\Phi$}\\
Switching the roles of the environment and system players leads to equally interesting questions about the effectiveness of coordination synthesis in collaboration. Specifically, in imperfect information games in unbounded and bounded scenes where the system player is the cops, will there always eventually be an \textit{arrest} (catching of a robber). The primary difference between this game set up and the previous game set up is that, if this game is realizable there will be an end, once the cops catch one of the robbers. Furthermore, this game removes the safe zones from the grid, so it is more akin to the traditional evasion-pursuit game of Cops and Robbers and adds slight rule modifications to avoid redundancy.
\begin{figure}[H]
\centering
\includegraphics[width=0.4\linewidth]{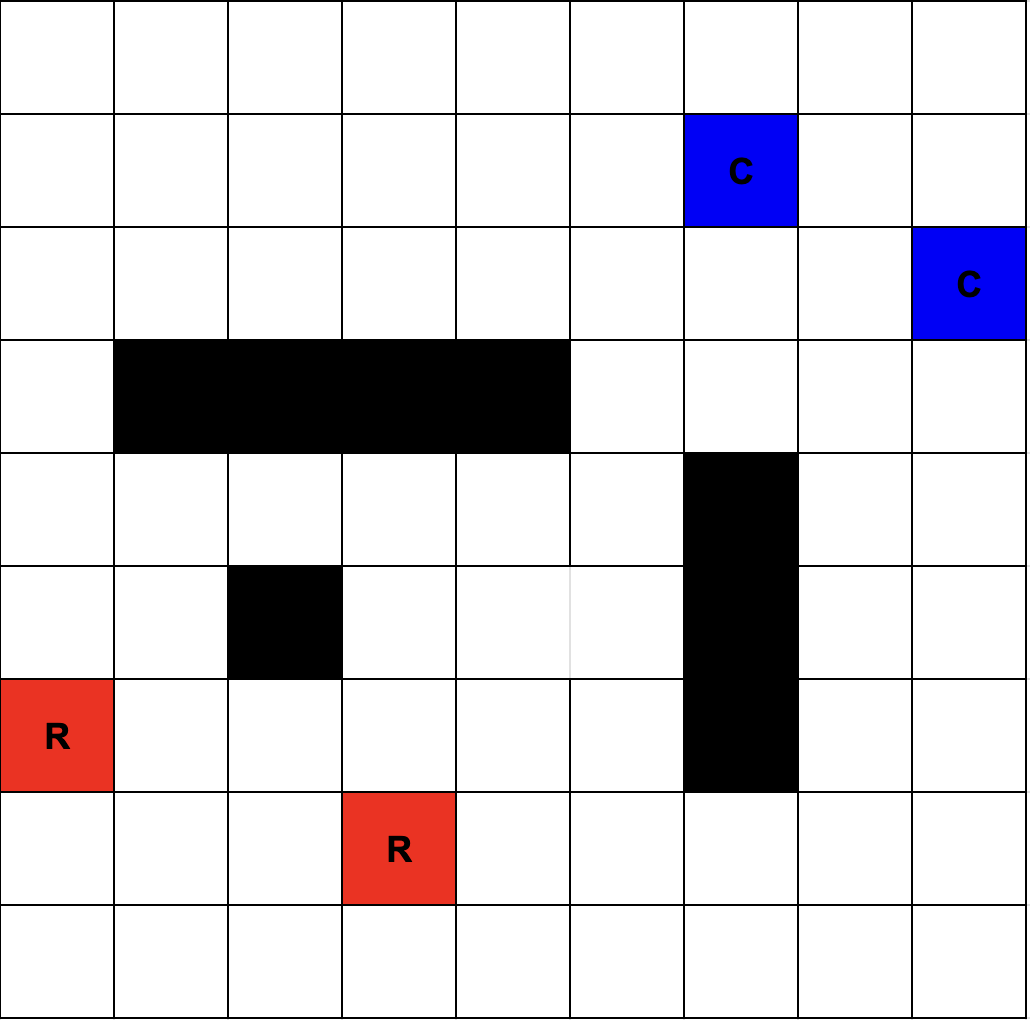}
\caption{Sample Initial Starting Position for Modified Game: Cops as System Players}
\end{figure}
\subsection{Imperfect Information Game: Finite And Infinite Grids}

The perfect information version is uninteresting because it is equivalent to a perfect information game where robbers act as system players. Similarly, without modification, the imperfect information game mirrors its robber analogue. To add interest, we propose two modifications—applied respectively to infinite and finite grid games:

\begin{enumerate}
    \item \textbf{Information Sharing:} At time \(T(i)\), cops within a Manhattan radius \(R\) share their entire radial view.
    \item \textbf{Map Memory:} Cops permanently remember the positions of fixed objects (e.g., walls) seen within their Zone of Interest (i.e., \(\forall i \in T\), all walls in \(ZI\) are remembered).
\end{enumerate}

To avoid state space explosion, \textit{Map Memory} is used only in finite grid games. The finite grid version is defined by the LTL specification
\begin{equation}
\begin{aligned}
\varphi = \Pi \land \mathbf{G}\Bigl( (\text{InZone} \land \text{WallSeen}) \to \mathbf{G}\,(\text{WallRemembered}) \Bigr)
\end{aligned}
\end{equation}
Conversely, to prevent finite grids from effectively becoming perfect information games due to crowding, \textit{Information Sharing} is applied only in infinite grid games. The infinite grid version is defined by the LTL specification
\begin{equation}
\begin{aligned}
\varphi = \Pi \land \mathbf{G}\Bigl( \text{CopsWithin}_R \to \text{ShareView} \Bigr)
\end{aligned}
\end{equation}

The inputs for these systems are slightly more complicated because they vary as function of \(T\) in the \textit{Map Memory} case or depending on what occupies the agent's radial view in the \textit{Information Sharing} case. While this complicates the construction of these systems, the rules for constructing the grid, the input types and outputs all are the same as in the imperfect information game for robbers.

An additional extension of map memory can also allow cops to remember positions of the robbers over time. 
There may then be opportunities for the cops to do observational learning of the strategy of the robbers at runtime.
We might assume that the robbers have a fixed, finite state strategy, in which case this becomes a sort of trace learning problem~\cite{zhang2024constrained}.
More generally, we might approach this with the goal of a probabilistic approximation of the robbers strategy.
This paradigm however does not fit well into the reactive synthesis model, which assume the environment (the robbers) are adversarial and can always play the ``best'' move to try to thwart the cops.

\section{Preliminary Thoughts: Coordination Synthesis And LTL\(t\) For Solving Cops and Robbers}
The goal of this research is to solve the Cops and Robbers game $\Phi$ and in solving this game be able to solve the set of more constrained problems defined from $\Phi$. Leveraging previous research in coordination synthesis with the express-ability of LTL\(t\), we aim to determine given an inputted grid layout and $R$ and $C$ cops and robbers is the reactive system realizable and generate strategies for the system players if realizable. A solution to this problem will also be mappable to a wide array of other solutions: multi agent warehouse logistics, hide and seek, capture the flag, etc... This system aims to be a useful tool in creating correct by construction models and potentially has applications in shielding and more advanced RL frameworks. 
\\
\\
Consider, we are given a set of environment agents \(e \in E\) and an LTL specification \(\mathbf{\Phi \land \psi}\). The goal is to construct a manager \(M\) such that agents can move without deadlock while continuously satisfying \(\mathbf{\Phi \land \psi}\). This manager must be able to communicate across the different agents and within each agent to ensure that the LTL specification's safety and liveness constraints are satisfied. We will represent the agents and the coordinator as CSP processes with a set of \textit{Public Actions} \(PA\) and \textit{Private Actions} \(PRA\). Given these environment agents, we can determine whether this problem is realizable if $\exists$ a CSP process \(M\) which represents the manager Bansal et al. \cite{coordination_synthesis}. The \(PA\)s represent the information that is shared throughout the system; this constitutes positional information such as the position of the robber and the cops. The \(PRA\)s represent more subtle aspects of each of the agents, such as choosing the position that minimizes distances to the safe zone, or the positional data of the safe zone; we plan to use \(PRA\)s as a way to ensure the liveness constraints of the system.
\\
\\
However, applying this formulation to our modified Cops and Robbers game is prohibitively difficult due to two factors: first, the state space is huge and constantly changing because the availability value of cells is updated as the robbers and cops move around the grid; second, the transition system grows very large, increasing by a factor of 8 with each additional agent. 
\\
\\
LTL\(t\) simplifies an agent's transition system by replacing the set of transitions 
\[
t_i \in \{t_1, t_2, \dots, t_8\}
\]
with a coordinate constraint:
\[
X - 1 < X < X + 1 \quad \text{or} \quad Y - 1 < Y < Y + 1.
\]
 In addition, we reduce the input space by representing the positions of the cops directly, rather than using boolean values for each cell. This approach minimizes the amount of information that needs to be tracked, as the system only needs to remember the current positions of the cops, the locations of safe zones and walls, and the positions of the robbers. Using real-valued coordinates eliminates the need to maintain memory for open spaces, thereby reducing the overall complexity of the problem further. Each player represents a \(e_i\) in the coordination synthesis problem, and the LTL\(t\) formulation can be re-expressed as a reactive program Rodriguez et al.\cite{Boolean_Abstractions,Reactive_Synthesis_Refs}. We leverage this, along with the ability to extract real values from the boolean abstractions, to perform coordination synthesis on these multi-agent systems, ensuring realizability and then the construction of a manager controller \(M\) which can then extract a valid strategy for each of the robbers.

\bibliographystyle{IEEEtran} 
\bibliography{refs}

\begin{thebibliography}{10}
\providecommand{\url}[1]{#1}
\csname url@samestyle\endcsname
\providecommand{\newblock}{\relax}
\providecommand{\bibinfo}[2]{#2}
\providecommand{\BIBentrySTDinterwordspacing}{\spaceskip=0pt\relax}
\providecommand{\BIBentryALTinterwordstretchfactor}{4}
\providecommand{\BIBentryALTinterwordspacing}{\spaceskip=\fontdimen2\font plus
\BIBentryALTinterwordstretchfactor\fontdimen3\font minus
  \fontdimen4\font\relax}
\providecommand{\BIBforeignlanguage}[2]{{%
\expandafter\ifx\csname l@#1\endcsname\relax
\typeout{** WARNING: IEEEtran.bst: No hyphenation pattern has been}%
\typeout{** loaded for the language `#1'. Using the pattern for}%
\typeout{** the default language instead.}%
\else
\language=\csname l@#1\endcsname
\fi
#2}}
\providecommand{\BIBdecl}{\relax}
\BIBdecl

\bibitem{finkbeiner2016synthesis}
B.~Finkbeiner, ``Synthesis of reactive systems,'' in \emph{Dependable Software
  Systems Engineering}.\hskip 1em plus 0.5em minus 0.4em\relax IOS Press, 2016,
  pp. 72--98.

\bibitem{pnueli1977temporal}
A.~Pnueli, ``The temporal logic of programs,'' in \emph{18th annual symposium
  on foundations of computer science (sfcs 1977)}.\hskip 1em plus 0.5em minus
  0.4em\relax ieee, 1977, pp. 46--57.

\bibitem{coordination_synthesis}
\BIBentryALTinterwordspacing
S.~Bansal, K.~S. Namjoshi, and Y.~Sa'ar, ``Synthesis of coordination programs
  from linear temporal specifications,'' \emph{Proceedings of the ACM on
  Programming Languages}, vol.~4, no. POPL, 2020. [Online]. Available:
  \url{https://dl.acm.org/doi/pdf/10.1145/3371122}
\BIBentrySTDinterwordspacing

\bibitem{Reactive_Synthesis_Refs}
A.~Rodriguez, F.~Gorostiaga, and C.~Sanchez, ``Predictable and performant
  reactive synthesis modulo theories via functional synthesis,'' in
  \emph{Automated Technology for Verification and Analysis}, S.~Akshay,
  A.~Niemetz, and S.~Sankaranarayanan, Eds.\hskip 1em plus 0.5em minus
  0.4em\relax Cham: Springer Nature Switzerland, 2025, pp. 28--50.

\bibitem{finkbeiner2019temporal}
B.~Finkbeiner, F.~Klein, R.~Piskac, and M.~Santolucito, ``Temporal stream
  logic: Synthesis beyond the bools,'' in \emph{International Conference on
  Computer Aided Verification}.\hskip 1em plus 0.5em minus 0.4em\relax
  Springer, 2019, pp. 609--629.

\bibitem{choi2022can}
W.~Choi, B.~Finkbeiner, R.~Piskac, and M.~Santolucito, ``Can reactive synthesis
  and syntax-guided synthesis be friends?'' in \emph{Proceedings of the 43rd
  ACM SIGPLAN International Conference on Programming Language Design and
  Implementation}, 2022, pp. 229--243.

\bibitem{nowakowski1983vertex}
R.~Nowakowski and P.~Winkler, ``Vertex-to-vertex pursuit in a graph,''
  \emph{Discrete Mathematics}, vol.~43, no. 2-3, pp. 235--239, 1983.

\bibitem{syntcomp}
S.~Jacobs, G.~A. P{\'e}rez, R.~Abraham, V.~Bruyere, M.~Cadilhac, M.~Colange,
  C.~Delfosse, T.~van Dijk, A.~Duret-Lutz, P.~Faymonville \emph{et~al.}, ``The
  reactive synthesis competition (syntcomp): 2018--2021,'' \emph{International
  journal on software tools for technology transfer}, vol.~26, no.~5, pp.
  551--567, 2024.

\bibitem{Boolean_Abstractions}
A.~Rodriguez and C.~Sanchez, ``Boolean abstractions for realizability modulo
  theories,'' in \emph{Computer Aided Verification}, C.~Enea and A.~Lal,
  Eds.\hskip 1em plus 0.5em minus 0.4em\relax Cham: Springer Nature
  Switzerland, 2023, pp. 305--328.

\bibitem{MultiRobotPusuitEvasion}
W.~Li, W.~Yan, H.~Shi, S.~Li, and Y.~Zhou, ``Multi-robot cooperative
  pursuit-evasion control: A deep-reinforcement learning approach based on
  prioritized experience replay,'' in \emph{Proceedings of the 2024 8th
  International Conference on Control Engineering and Artificial Intelligence},
  Shanghai, China, 2024, pp. 120--127.

\bibitem{xu2024approximate}
Z.~Xu, D.~Yu, Y.-J. Liu, and Z.~Wang, ``Approximate optimal strategy for
  multiagent system pursuit--evasion game,'' \emph{IEEE Systems Journal}, 2024.

\bibitem{crawford2024k}
N.~Crawford and V.~I. Chenoweth, ``{$k$-Hyperopic Cops and Robber},'' 2024,
  arXiv preprint \texttt{arXiv:2410.17678}.

\bibitem{erlebach2024cop}
T.~Erlebach, N.~Morawietz, J.~T. Spooner, and P.~Wolf, ``A cop and robber game
  on edge-periodic temporal graphs,'' \emph{Journal of Computer and System
  Sciences}, vol. 144, p. 103534, 2024.

\bibitem{katis2018validity}
\BIBentryALTinterwordspacing
A.~Katis, G.~Fedyukovich, H.~Guo, A.~Gacek, J.~Backes, A.~Gurfinkel, and M.~W.
  Whalen, ``Validity-guided synthesis of reactive systems from assume-guarantee
  contracts,'' in \emph{Proc. of the 24th Int'l Conf. on Tools and Algorithms
  for the Construction and Analysis of Systems, (TACAS'18), Part {II}}, ser.
  LNCS, vol. 10806.\hskip 1em plus 0.5em minus 0.4em\relax Springer, 2018, pp.
  176--193. [Online]. Available:
  \url{https://doi.org/10.1007/978-3-319-89963-3\_10}
\BIBentrySTDinterwordspacing

\bibitem{samuel21gensys}
\BIBentryALTinterwordspacing
S.~Samuel, D.~D'Souza, and R.~Komondoor, ``Gensys: a scalable fixed-point
  engine for maximal controller synthesis over infinite state spaces,'' in
  \emph{Proc. of the 29th {ACM} Joint European Software Engineering Conference
  and Symposium on the Foundations of Software Engineering ({ESEC/FSE}
  '21)}.\hskip 1em plus 0.5em minus 0.4em\relax {ACM}, 2021, pp. 1585--1589.
  [Online]. Available: \url{https://doi.org/10.1145/3468264.3473126}
\BIBentrySTDinterwordspacing

\bibitem{finkbeiner21temporal}
\BIBentryALTinterwordspacing
B.~Finkbeiner, P.~Heim, and N.~Passing, ``Temporal stream logic modulo
  theories,'' in \emph{Proc. of the 25th Int'l Conf. on Foundations of Software
  Science and Computation Structures (FOSSACS'22)}, ser. LNCS, vol.
  13242.\hskip 1em plus 0.5em minus 0.4em\relax Springer, 2022, pp. 325--346.
  [Online]. Available: \url{https://doi.org/10.1007/978-3-030-99253-8\_17}
\BIBentrySTDinterwordspacing

\bibitem{maderbacher2021reactive}
\BIBentryALTinterwordspacing
B.~Maderbacher and R.~Bloem, ``Reactive synthesis modulo theories using
  abstraction refinement,'' in \emph{Proc. of the 22nd Int'l Conf. on Formal
  Methods in Computer-Aided Design, (FMCAD'22)}.\hskip 1em plus 0.5em minus
  0.4em\relax {IEEE}, 2022, pp. 315--324. [Online]. Available:
  \url{https://doi.org/10.34727/2022/isbn.978-3-85448-053-2\_38}
\BIBentrySTDinterwordspacing

\bibitem{heim24solving}
\BIBentryALTinterwordspacing
P.~Heim and R.~Dimitrova, ``Solving infinite-state games via acceleration,''
  \emph{Proc. {ACM} Program. Lang.}, vol.~8, no. {POPL}, pp. 1696--1726, 2024.
  [Online]. Available: \url{https://doi.org/10.1145/3632899}
\BIBentrySTDinterwordspacing

\bibitem{schmuck24localized}
\BIBentryALTinterwordspacing
A.~Schmuck, P.~Heim, R.~Dimitrova, and S.~P. Nayak, ``Localized attractor
  computations for infinite-state games,'' in \emph{Proc. of the 36th Int'l
  Conf. on Computer Aided Verification (CAV'24), Part {III}}, ser. LNCS, vol.
  14683.\hskip 1em plus 0.5em minus 0.4em\relax Springer, 2024, pp. 135--158.
  [Online]. Available: \url{https://doi.org/10.1007/978-3-031-65633-0\_7}
\BIBentrySTDinterwordspacing

\bibitem{heim25efficient}
\BIBentryALTinterwordspacing
P.~Heim and R.~Dimitrova, ``Translation of temporal logic for efficient
  infinite-state reactive synthesis,'' \emph{Proc. {ACM} Program. Lang.},
  vol.~9, no. {POPL}, 2025. [Online]. Available:
  \url{https://doi.org/10.1145/3704888}
\BIBentrySTDinterwordspacing

\bibitem{geatti22linear}
\BIBentryALTinterwordspacing
L.~Geatti, A.~Gianola, and N.~Gigante, ``Linear temporal logic modulo theories
  over finite traces,'' in \emph{Proc. of the 31st Int'l Joint Conf. on
  Artificial Intelligence, ({IJCAI}'22)}.\hskip 1em plus 0.5em minus
  0.4em\relax ijcai.org, 2022, pp. 2641--2647. [Online]. Available:
  \url{https://doi.org/10.24963/ijcai.2022/366}
\BIBentrySTDinterwordspacing

\bibitem{geatti23decidable}
\BIBentryALTinterwordspacing
L.~Geatti, A.~Gianola, N.~Gigante, and S.~Winkler, ``Decidable fragments of
  {LTL${_f}$} modulo theories,'' in \emph{Proc. of the 26th European Conference
  on Artificial Intelligence (ECAI'23)}, ser. Frontiers in Artificial
  Intelligence and Applications, vol. 372.\hskip 1em plus 0.5em minus
  0.4em\relax {IOS} Press, 2023, pp. 811--818. [Online]. Available:
  \url{https://doi.org/10.3233/FAIA230348}
\BIBentrySTDinterwordspacing

\bibitem{rodriguez24realizability}
A.~Rodr\'iguez and C.~S\'{a}nchez, ``Realizability modulo theories,''
  \emph{Journal of Logical and Algebraic Methods in Programming}, vol. 140, p.
  100971, 2024.

\bibitem{rodriguez24adaptive}
------, ``{Adaptive Reactive Synthesis for {LTL} and {LTLf} Modulo Theories},''
  in \emph{Proc. of the 38th AAAI Conf. on Artificial Intelligence ({AAAI}
  2024)}.\hskip 1em plus 0.5em minus 0.4em\relax {AAAI} Press, 2024, pp.
  10\,679--10\,686.

\bibitem{rodriguez25shield}
A.~Rodr\'iguez, G.~Amir, D.~Corsi, C.~S\'{a}nchez, and G.~Katz, ``Shield
  synthesis for {LTL} modulo theories,'' in \emph{Proc. of the 39th AAAI Conf.
  on Artificial Intelligence (AAAI'25)}.\hskip 1em plus 0.5em minus 0.4em\relax
  {AAAI} Press, 2025.

\bibitem{corsi24verification}
D.~Corsi, G.~Amir, A.~Rodr{\'{\i}}guez, G.~Katz, C.~S{\'{a}}nchez, and R.~Fox,
  ``Verification-guided shielding for deep reinforcement learning,''
  \emph{{RLJ}}, vol.~4, pp. 1759--1780, 2024.

\bibitem{rutledge2023controller}
K.~Rutledge, Y.~Mei, and N.~Ozay, ``Controller synthesis for unknown-mode
  linear systems with an epistemic variant of ltl,'' in \emph{Proceedings of
  the 2023 American Control Conference (ACC)}, 2023, pp. 3508--3515.

\bibitem{zhang2024constrained}
C.~Zhang, P.~Kapoor, I.~Dardik, L.~Cui, R.~Meira-Goes, D.~Garlan, and E.~Kang,
  ``Constrained ltl specification learning from examples,'' \emph{arXiv
  preprint arXiv:2412.02905}, 2024.

\end{thebibliography}

\end{document}